%% file: ettorezaffaroni.tex
\documentclass{moriond}
\usepackage{siunitx}
\usepackage{caption}
\usepackage{subcaption}

\newcommand{\Photo}{}

\begin{document}
\vspace*{4cm}
\title{Results from SND@LHC}

\author{ E.~Zaffaroni, on behalf of the SND@LHC collaboration }

\address{Institute of Physics, Ecole Polytechnique Fédérale de Lausanne (EPFL),\\
 Lausanne, Switzerland}

\maketitle\abstracts{
  SND@LHC is a compact and stand-alone experiment to perform measurements with neutrinos produced at the LHC in a hitherto unexplored pseudo-rapidity region of $7.2 < \eta < 8.4$, complementary to all the other experiments at the LHC.
  The experiment is located 480 m downstream of IP1 in the unused TI18 tunnel.
  The detector is composed of a hybrid system based on an 800 kg target mass of tungsten plates, interleaved with emulsion and electronic trackers, followed downstream by a calorimeter and a muon system.
  The configuration allows efficiently distinguishing between all three neutrino flavours, opening a unique opportunity to probe physics of heavy flavour production at the LHC in the region that is not accessible to ATLAS, CMS and LHCb.
  SND@LHC has started taking data in 2022 and in this work the direct observation of muon neutrino interaction in the detector is reported.
}

\input{sections/introduction.tex}
\input{sections/detector.tex}

\input{sections/data.tex}
\input{sections/analysis.tex}

\input{sections/background.tex}

\input{sections/significance.tex}
\input{sections/conclusions.tex}



\section*{References}
\bibliography{bibliography}

\end{document}

%% file: sections/introduction.tex
\section{Introduction}
\label{sec:intro}

The use of LHC as a neutrino factory was first envisaged about 30 years ago~\cite{DeRujula:1984pg,DeRujula:1992sn,Vannucci:253670}, in particular for the then undiscovered $\nu_{\tau}$~\cite{Jarlskog:215298}.
The idea suggested a detector intercepting the very forward flux (${\eta > 7}$) of neutrinos (about 5\% have $\tau$ flavour) from $b$ and $c$ decays.


Neutrinos allow precise tests of the Standard Model (SM)~\cite{brock,conrad,formaggio,delellis} and are a probe for 
new physics~\cite{marfatia,arguelles}. Measurements of the neutrino cross section in the last decades were mainly performed at low energies. The region between \SI{350}{GeV} and \SI{10}{TeV} is currently unexplored~\cite{bustamante}. 

SND@LHC~\cite{Ahdida:2750060} was designed to perform measurements with high-energy neutrinos (\SI{100}{GeV} to a few TeV) 
produced at the LHC in the pseudo-rapidity region $7.2<\eta<8.4$. It is a compact, standalone experiment located in the TI18 transfer tunnel (480~m downstream of the ATLAS interaction point) and it allows for the identification of all three flavours of neutrino interactions with high efficiency. 

The detector was installed in TI18 in 2021 during the Long Shutdown~2 and it has started to collect data since the beginning of the LHC Run~3 in April 2022. The experiment will run throughout the whole Run~3 and it is expected to collect 250 fb$^{-1}$ of data in 2022--25, corresponding to a total of two thousand expected high-energy neutrino interactions of all flavours in the detector target.


In this work, we report the detection of $\nu_\mu$ charged current (CC) interactions using the data that was taken by the electronic detectors of the SND@LHC experiment in 2022.

%% file: sections/detector.tex
\section{Detector}
\label{sec:detector}

\begin{figure}[h]
  \centering
  \includegraphics[width=0.95\textwidth]{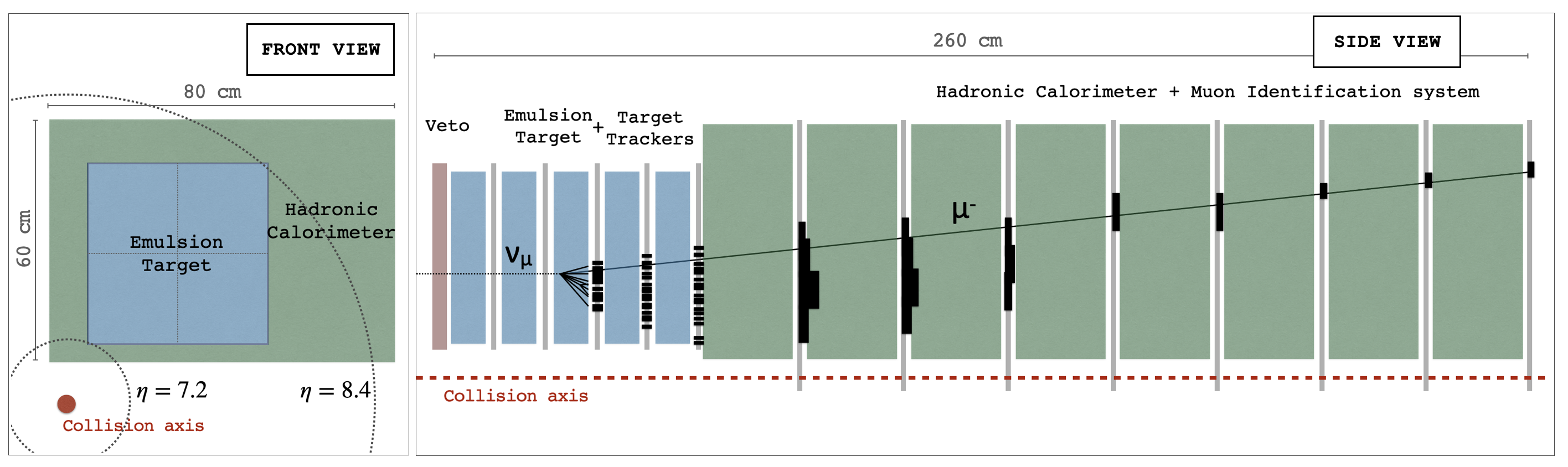}
  \caption{Layout of the SND@LHC detector.}
  \label{fig:detector}
  \end{figure}

The SND@LHC detector consists of a hybrid system with a $\sim$830 kg target made of tungsten plates interleaved with nuclear emulsion and electronic trackers, followed by a hadronic calorimeter and a muon identification system (see Figure~\ref{fig:detector}). The electronic detectors provide the time stamp of the neutrino interaction, preselect the interaction region, identify muons and measure the electromagnetic and hadronic energy, while the emulsion detectors provide excellent vertex reconstruction.

The detector consists of three parts: the Veto, the target and electromagnetic calorimeter, the hadronic calorimeter and muon systems. 

The Veto system is located upstream of the target region and comprises two parallel planes of stacked scintillating bars read out by SiPMs.
This system is used to identify muons and other charged particles entering the detector from the IP1 direction.

The target section contains five walls.
Each wall consists of four units of Emulsion Cloud Chambers (ECC) and is followed by a Scintillating Fibre  tracker (SciFi) station.
Each ECC module is a sequence of 60 emulsion films, $19.2 \times 19.2$ cm$^2$, interleaved with 59 tungsten plates, 1~mm thick. 
Its weight is approximately 41.5~kg, adding up to about 830~kg for the total target mass.


Each SciFi station consists of five $40 \times 40$ cm$^2$ $x$-$y$ planes of staggered  scintillating fibres with a diameter of \SI{250}{\micro m}.
The single particle spatial resolution is of order of \SI{\sim150}{\micro m} and the time resolution for a particle crossing a $x$ and $y$ plane \SI{\sim 250}{ps}.

The muon system and hadronic calorimeter consists of two parts: upstream (US), the first five stations, and downstream (DS), the last three stations. In combination with SciFi, it acts as a coarse sampling calorimeter ($\sim~9.5~\lambda_{\rm{int}}$), providing the energy measurement of hadronic jets.
Each US station consists of 10 stacked horizontal scintillator bars, a DS station consists of two layers of thin bars, arranged horizontally and vertically, allowing for a spatial resolution less than 1~cm and allowing to isolate muons.
The eight scintillating planes are interleaved with 20~cm thick iron blocks. 

All signals exceeding preset thresholds are read out by the front-end electronics and clustered in time to form events.
A software noise filter is applied to the events online, resulting in negligible detector deadtime or loss in signal efficiency.

%% file: sections/data.tex
\section{Dataset and simulated samples}
\label{sec:data}

The data collected during 2022, with {\it pp} collisions with a center of mass energies of \SI{13.6}{TeV}, has been analysed.
The integrated luminosity estimated by the ATLAS collaboration~\cite{ATLAS:2022hro} to have been delivered at IP1 during this period was \SI{39.0}{fb^{-1}}, of which \SI{37.6}{fb^{-1}} were recorded, corresponding to a detector uptime of 96\%.

We describe the analysis developed for the first observation of $\nu_\mu$ charged-current interactions from LHC collisions.
This analysis is conducted solely using the data from the electronic detectors, as information from the emulsion target is being processed.

In SND@LHC the dominant Charged Current process occurring for $\nu_\mu$s is the Deep Inelastic Scattering (CCDIS), given the high energy of neutrinos within the detector acceptance~\cite{Ahdida:2750060}.
The signature of these interactions includes an isolated muon track in the Muon Identification system, pointing back to a shower which is developing in the SciFi and hadronic calorimeter.
In Figure~\ref{fig:detector} the distinctive topology of $\nu_\mu$ CCDIS interactions in the SND@LHC detector is shown.

Neutrino production in proton-proton collisions at the LHC is simulated with the \textsc{FLUKA} Monte Carlo code~\cite{Fluka2}.
\textsc{DPMJET3} (Dual Parton Model, including charm)~\cite{Roesler_2001,DPMJET}  is used for the $pp$ event generation, and \textsc{FLUKA} performs the particle propagation towards the SND@LHC detector with the help of a detailed simulation of LHC accelerator elements.
\textsc{FLUKA} also takes care of simulating the production of neutrinos from decays of long-lived products of the $pp$ collisions and of particles produced in re-interactions with the surrounding material.
\textsc{Genie}~\cite{cite:GENIE} is then used to simulate neutrino interactions with the SND@LHC detector material.

Given the total mass of the Tungsten target during the 2022 run (\SI{\sim 800}{kg}), about $160 \pm 38$ $\nu_\mu$ CCDIS interactions are expected in the full target in the analysed data set.

%% file: sections/analysis.tex
\section{Analysis}
Observing the rare neutrino signal over the prevailing background implies adopting a selection with strong rejection power, designed to yield a set of clean events.

The signal selection proceeds in two steps. 
The first step aims at identifying events happening in a fiducial region of the target, while rejecting backgrounds entering from the front and edges of the detector. It selects events induced by a neutral-like particle located in the $3^{\mathrm{rd}}$ or $4^{\mathrm{th}}$ target wall. The exclusion of events starting in the two most upstream target walls enhances the rejection power for muon-induced backgrounds, while excluding events starting in the most downstream wall ensures the neutrino-induced showers are sampled by at least two SciFi planes. The average SciFi channel and DS bar number are used to discard events with hits at the edges of detectors' sensitive areas, resulting in a fiducial cross-sectional area in the XY plane of 25 $\times$ 26 cm$^{2}$.  The efficiency of  fiducial region cuts on simulated neutrino interactions in the target is $7.5 \times 10^{-2}$.

\begin{figure}[h]
\centering
\includegraphics[width=0.5\textwidth]{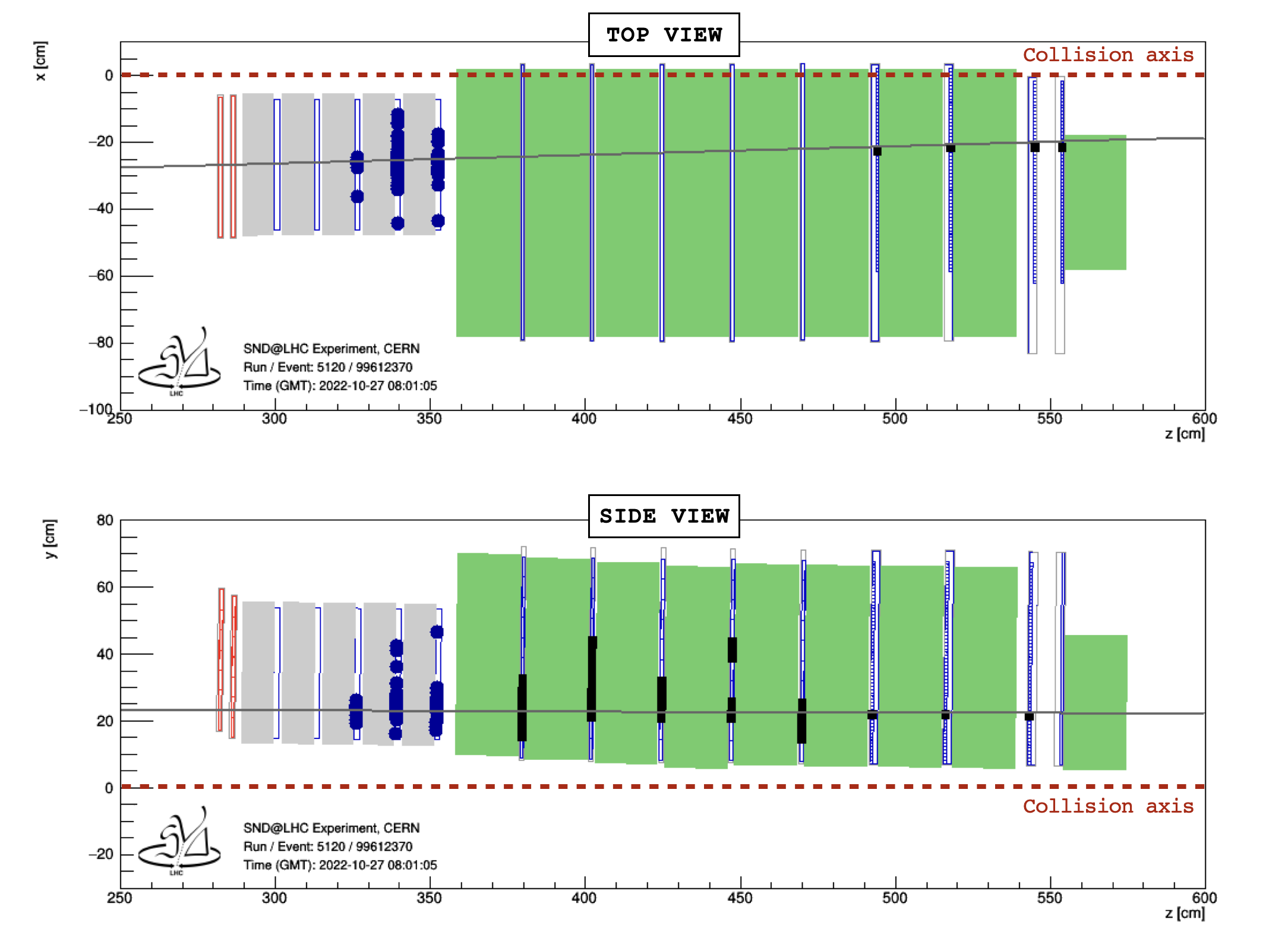}
\caption{Display of a $\nu_\mu$ CC candidate event. Hits in the SciFi and hadronic calorimeter are shown as blue markers and black bars, respectively, and the line represents the reconstructed muon track.}
\label{fig:candidate}
\end{figure}

The second step of the selection aims at selecting signal-like signature patterns using a cut-based procedure. $\nu_\mu$ CCDIS interactions are associated to a large hadronic activity in the calorimetric system, with a clean outgoing muon track reconstructed in the Muon Identification system, and hit time distribution consistent with an event originating from the IP1 direction. The muon track is defined by a set of Muon Identification hits in a straight-line pattern spanning at least three detector planes in both ZX and ZY views. Events with a large number of hits in the Muon Identification system are rejected to ensure cleanly reconstructed tracks.

The achieved reduction factor on the data for the total selection (fiducial and neutrino identification cuts) amounts to $1.0\times 10^{9}$, while the overall efficiency on the  $\nu_\mu$ CCDIS Monte Carlo sample is $2.9\times 10^{-2}$. The event selection increases the signal-to-noise ratio by 10 orders of magnitude. 

As a result of the full selection, 8 $\nu_\mu$ CCDIS candidates are identified, while 4.6 are expected. The contribution of other neutrino flavours and neutral current interactions to the selected sample is less than 1\% of the expected $\nu_\mu$ CCDIS yield.
One of the selected candidates is shown in Figure~\ref{fig:candidate}.

%% file: sections/background.tex
\section{Background}
\label{sec:background}
Muons reaching the detector location are the main source of background for the neutrino search.
They can either enter in the fiducial volume without being vetoed and generate showers via bremsstrahlung or deep inelastic scattering, or interact in the surrounding material and produce neutral particles that can then mimic neutrino interactions in the target.

The estimate of the penetrating muon background based on the expected flux in the fiducial volume and on the inefficiency of detector planes used as veto: the Veto system and the two upstream Scifi planes.

The simulation of the muon flux at the detector location is provided by the CERN EN-STI team.
Proton-proton interactions are simulated with \textsc{FLUKA}~\cite{Fluka2} and charged $\pi$'s and K's from IP1 are transported along the LHC straight section until their decay. 

The total number of muons integrated in 37.6 fb$^{-1}$ amounts to $4.7\times10^8$.

The inefficiency of the Veto planes is estimated from data by using good quality tracks reconstructed in the SciFi detector and confirmed with a track segment in the DS detector.
The tracks are extrapolated to the Veto fiducial volume.
All tracks are identified as muons due to the large interaction length; tracks entering the detector from the downstream end are excluded by timing measurements.
The overall Veto system inefficiency during the 2022 run amounts to $4.4 \times 10^{-4}$. 

\begin{figure}
  \centering
  \begin{subfigure}[b]{0.4\textwidth}
      \centering
      \includegraphics[width=\textwidth]{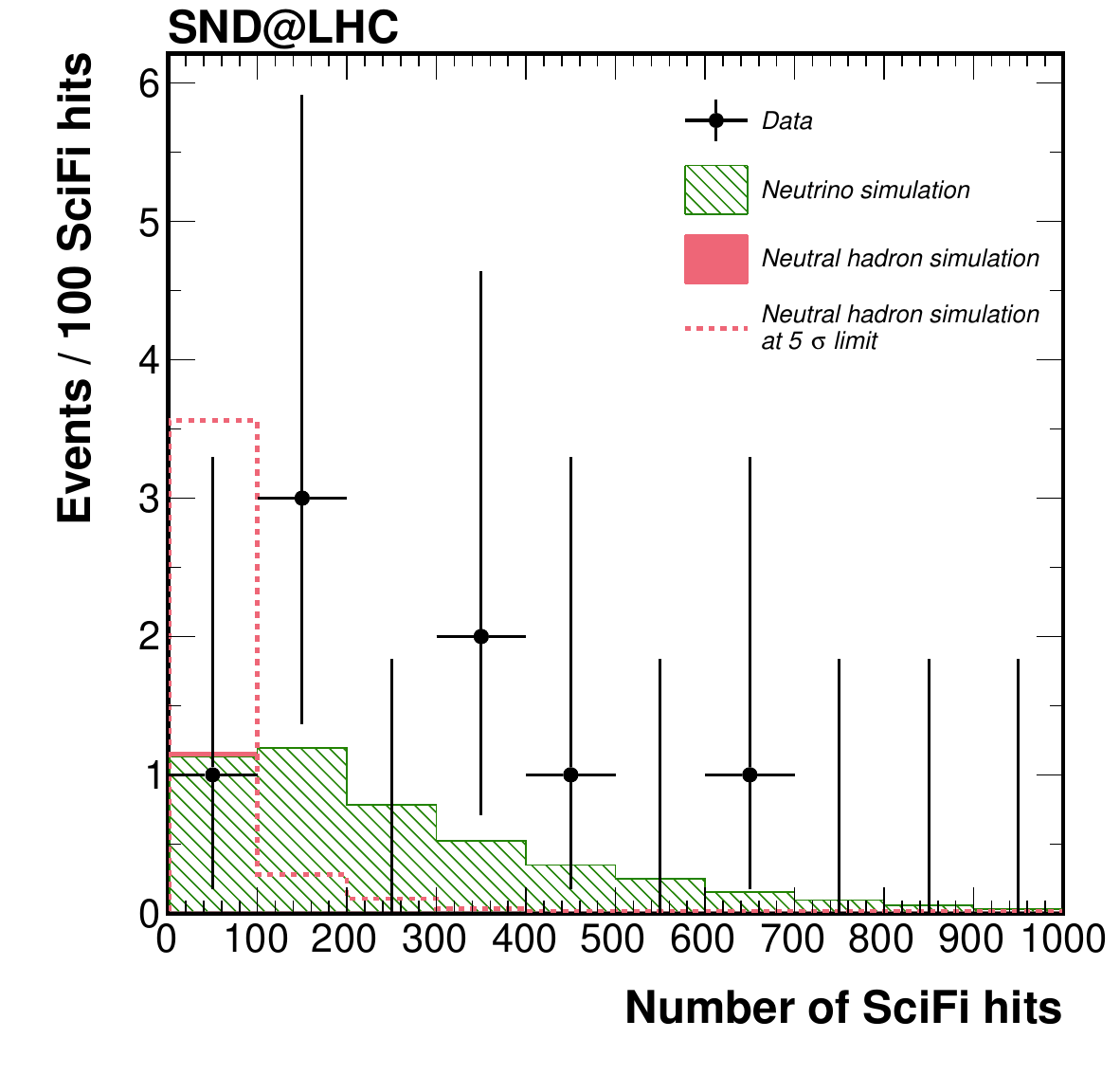}
      \caption{}
      \label{fig:scifi_hits}
  \end{subfigure}
  \hfill
  \begin{subfigure}[b]{0.4\textwidth}
      \centering
      \includegraphics[width=\textwidth]{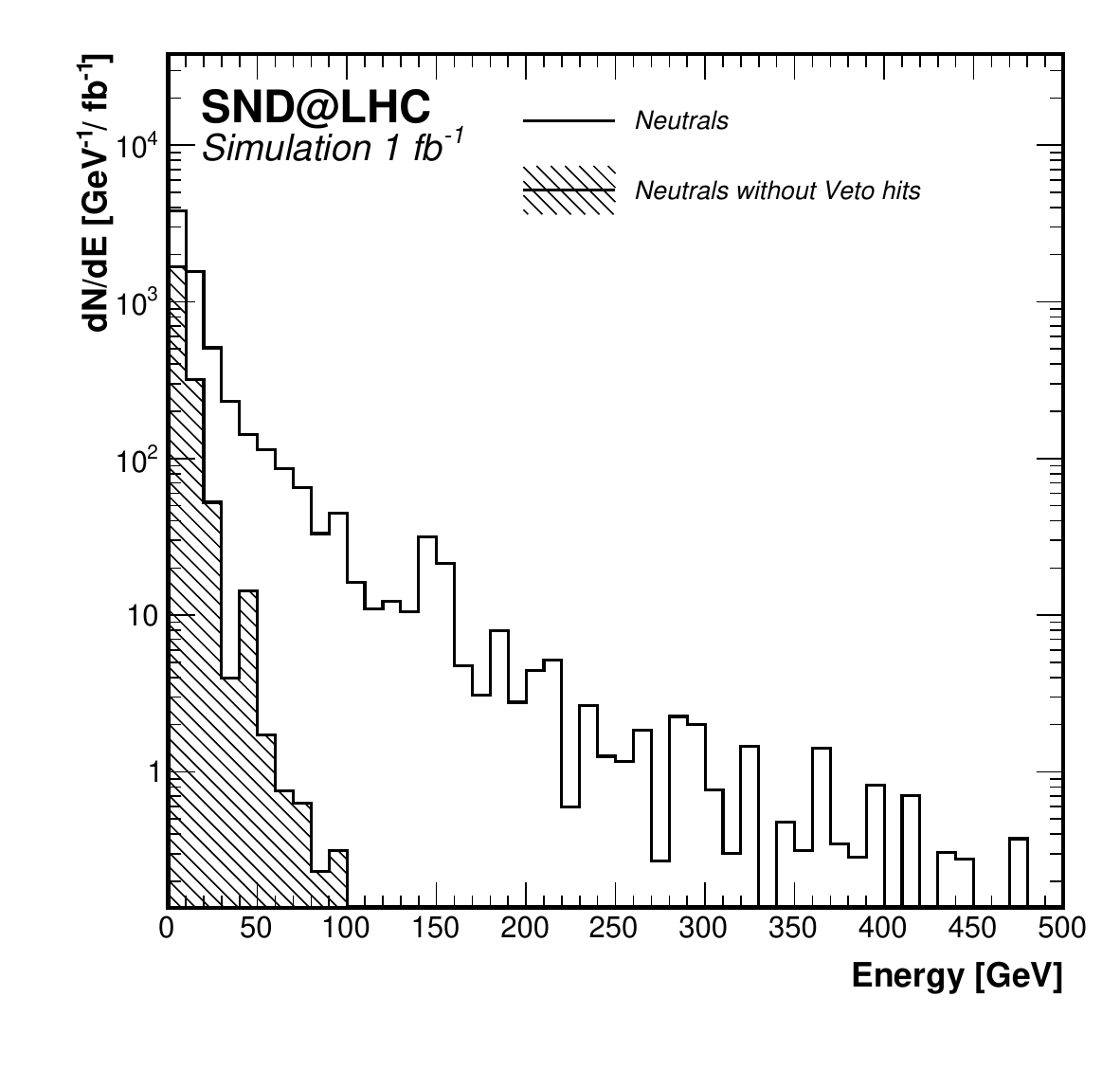}
      \caption{}
      \label{fig:neutral_energy_spectrum}
  \end{subfigure}
     \caption{ (a) Distribution of SciFi hits for candidate events with the expectation from neutrino signal and neutral hadron background overlayed. The dashed line represents the background-only hypothesis scaled up to a 5 $\sigma$ statistical deviation from the nominal expectation. (b) Energy spectrum of neutral hadrons produced by muon interactions in the rock and concrete entering the SND@LHC acceptance. The dashed line shows the spectrum after rejecting events with hits in the Veto detector.}
     \label{fig:plots}
\end{figure}



The SciFi detector inefficiency is estimated with a similar method used for the Veto, using reconstructed SciFi tracks confirmed with a DS track and hits in the Veto system, and it amounts to $1.1\times 10^{-4}$ per station.
The combined inefficiency of the Veto system and the two most upstream SciFi planes is therefore $5.3\times10^{-12}$, thus making the background induced by muons entering the fiducial volume negligible.

Neutral particles (mainly neutrons and $K_L$'s) originating from primary muons interacting in rock and concrete in front of the detector can potentially mimic a neutrino interaction since they do not leave any incoming trace in the electronic detectors, and can create a vertex in the target with a DS track produced by punch-through or decay-in-flight $\pi^{\pm}$ and $K^{\pm}$.
Although they are mainly rejected due to accompanying charged particles originating from the primary muon interaction, they constitute the main background source for the neutrino search. The energy spectrum of neutral hadrons entering the SND@LHC detector is shown in Figure~\ref{fig:neutral_energy_spectrum}, where the suppression achieved by rejecting events in which accompanying charged particles produce hits in the Veto detector is also shown.
The vast majority of events with neutral hadrons above around 20 GeV are rejected by Veto hits produced by the scattered muon.

\textsc{Pythia v6.4} was used to simulate photo nuclear interactions of $\mu^+$ and $\mu^-$ on protons or neutrons at rest using the muon spectrum expected at the detector location.
These events are placed along the muon flight direction according to the material density, and the secondary particles are transported by \textsc{GEANT4} in the detector surroundings.
Neutral particles induced by muon DIS make interactions in the rock and concrete and only a small fraction of the particles leaves the tunnel wall and enters the SND@LHC detector. 

To estimate the yield of neutral particles passing the event selection criteria, we simulate the highest energy neutral hadrons entering the SND@LHC target region in a given muon DIS interaction using \textsc{Geant4}. The events are simulated with energies ranging from $[5,\,200]\,$GeV and uniformly distributed across the front face of the SND@LHC target.
As shown in Figure~\ref{fig:neutral_energy_spectrum}, the rate of neutral hadrons events with energies above 100 GeV is heavily suppressed by using the Veto to tag the accompanying charged particles (most often the scattered muon), and below 5 GeV the minimum ionizing particles resulting from the neutral hadron interactions do not have enough energy to produce a track exiting the downstream end of the SND@LHC detector.

As can be seen in Figure~\ref{fig:scifi_hits}, the lower energy of the neutral hadrons, compared to the neutrino signal, results in fewer hits in the SciFi detector.
We note that while such variables have not been used in the present analysis, they are powerful in discriminating neutrino events from the neutral hadron background.

The residual background yield results from the convolution of the selection efficiency with the yield of neutral hadrons in the acceptance and not accompanied by a charged track producing hits in the Veto detector. The background yield after the selection amounts to $(7.8 \pm 2.9) \times 10^{-2}$ and is dominated by neutrons and $K^{0}_{L}$s.


%% file: sections/significance.tex
\section{Significance evaluation}
\label{subsec:significance}

The significance of the observation of 8 candidates with an expected background yield of $7.8 \times 10^{-2}$ is addressed by considering the confidence in the exclusion of the null hypothesis (i.e. having no neutrino signal). 
The one-sided profile likelihood ratio test $\lambda(\mu)$ was also used as test statistic. The significance was evaluated by comparing $\lambda_{data}(\mu=0)$  with the sampling distribution of $\lambda(\mu=0)$. The likelihood, which includes Gaussian terms to account for the background uncertainties, is
\begin{displaymath}
L = \mathrm{Poisson}(n\,|\, \mu s + \beta)\, \mathrm{Gauss}(\beta\,|\,b,\sigma_b)
\end{displaymath}
where $\sigma_b$ is the background uncertainty and $\beta$ are the background parameters Gaussian modeled. The implementations of the method is based on RooStats, giving $p=1.0\times10^{-12}$, corresponding to a $7.03~\sigma$ exclusion.

%% file: sections/conclusions.tex
\section{Conclusions}
\label{sec:conclusions}

A search for high energy neutrino neutrinos originating from $pp$ collisions is presented using data taken by the electronic detectors from SND@LHC installed at the LHC in
2022. We observe 8 candidate events consistent
with $\nu_\mu$ CC interactions at the LHC.
Our muon-induced and neutral backgrounds for the analysed dataset amount to $(7.8 \pm 2.9) \times 10^{-2}$ events, which implies a $7.0~\sigma$ excess of $\nu_\mu$ CC signal events.